\documentclass[review]{elsarticle}
\usepackage{hyperref}
\hypersetup{hidelinks}
\usepackage[margin=1in]{geometry}
\usepackage{etoolbox}
\usepackage{subcaption}
\usepackage{booktabs}
\AtBeginEnvironment{thebibliography}{\interlinepenalty=10000}
\usepackage{gensymb} 

\journal{arXiv}

\begin{document}

\begin{frontmatter}

\title{Real-time accident detection and physiological signal monitoring to enhance motorbike safety and emergency response}

\author[1]{S. M. Kayser Mehbub Siam}

\author[1]{Khadiza Islam Sumaiya}

\author[1]{Md Rakib Al-Amin}

\author[1]{Tamim Hasan Turjo}

\author[2]{Ahsanul Islam}

\author[1]{A.H.M.A. Rahim}

\author[1,3]{Md Rakibul Hasan\corref{corr}}
\cortext[corr]{Corresponding author}
\ead{Rakibul.Hasan@curtin.edu.au}
\affiliation[1]{organization={BRAC University},
city={Dhaka},
country={Bangladesh}}

\affiliation[2]{organization={Khulna University of Engineering and Technology},
city={Khulna},
country={Bangladesh}}

\affiliation[3]{organization={Curtin University},
city={Perth WA},
country={Australia}}

\begin{abstract}
Rapid urbanization and improved living standards have led to a substantial increase in the number of vehicles on the road, consequently resulting in a rise in the frequency of accidents. Among these accidents, motorbike accidents pose a particularly high risk, often resulting in serious injuries or deaths. A significant number of these fatalities occur due to delayed or inadequate medical attention. To this end, we propose a novel automatic detection and notification system specifically designed for motorbike accidents. The proposed system comprises two key components: a detection system and a physiological signal monitoring system. The detection system is integrated into the helmet and consists of a microcontroller, accelerometer, GPS, GSM, and Wi-Fi modules. The physio-monitoring system incorporates a sensor for monitoring pulse rate and SpO$_{2}$ saturation. All collected data are presented on an LCD display and wirelessly transmitted to the detection system through the microcontroller of the physiological signal monitoring system. If the accelerometer readings consistently deviate from the specified threshold decided through extensive experimentation, the system identifies the event as an accident and transmits the victim’s information -- including the GPS location, pulse rate, and SpO$_{2}$ saturation rate -- to the designated emergency contacts. Preliminary results demonstrate the efficacy of the proposed system in accurately detecting motorbike accidents and promptly alerting emergency contacts. We firmly believe that the proposed system has the potential to significantly mitigate the risks associated with motorbike accidents and save lives.
\end{abstract}


\end{frontmatter}

\section{Introduction}
\begin{figure}[!htbp]
\centering
\includegraphics[width=0.8\textwidth]{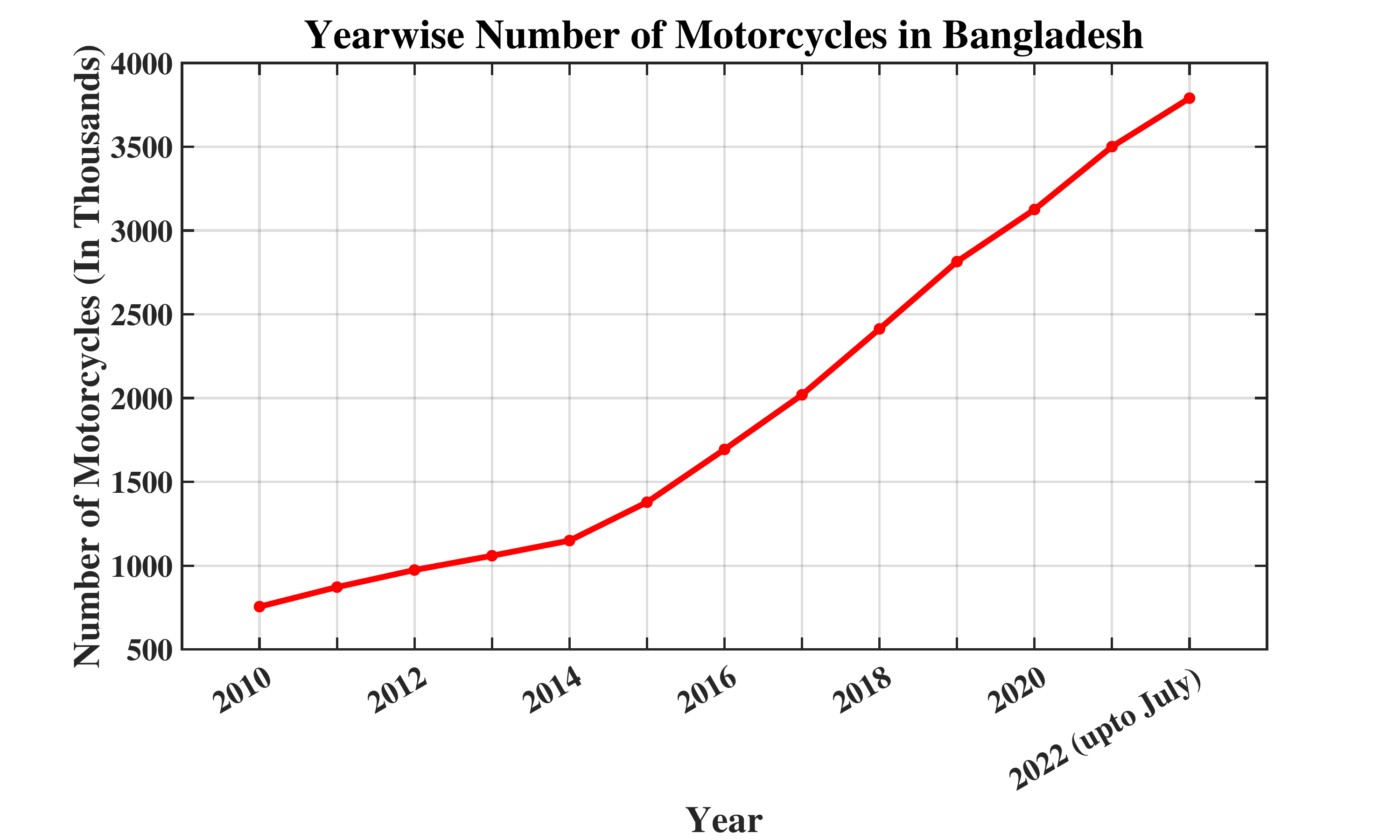}
\caption{Yearwise growth of motorbikes in Bangladesh \citep{rf1}.}
\label{ymb}
\end{figure}

Rapid urbanization has led to a staggering growth of civilization, which has further increased the standard of living. More and more people now can buy their own motor vehicles, and the number of motor vehicles on the streets is increasing day by day. Among all the motor vehicles that run on the streets, motorcycles are the most cost-efficient and affordable for most people. About 71\% of all registered vehicles in Bangladesh are motorcycles \citep{rf1}. Fig.~\ref{ymb} shows how rapidly the number of motorcycles has increased. In 10 years, it has gone from only a few hundred thousand to almost 3.7 million \citep{rf1}.

\begin{figure}[!ht]
\centering
\includegraphics[width=0.8\textwidth]{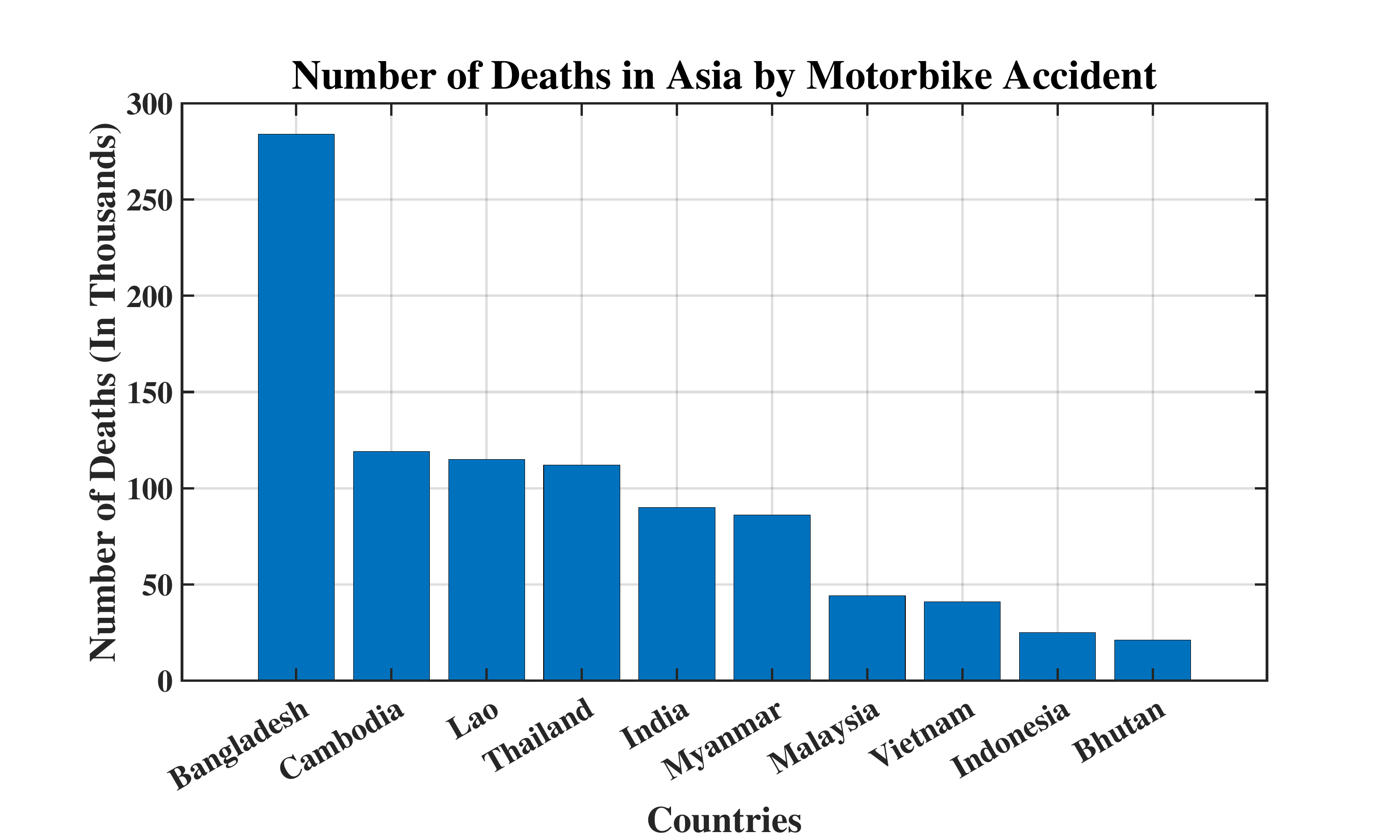}
\caption{Number of deaths by motorbikes in Asia \citep{rf2}.}
\label{mbac}
\end{figure}

Although motor vehicles are the best means of transportation nowadays, more motor vehicles on the streets mean a rise in automobile accidents as well.
An analysis carried out by the Accident Research Institute at the Bangladesh University of Engineering Technology revealed that the deaths from motorcycle accidents are 74\% in rural areas and 45\% on national highways \citep{rf2}. Fig.~\ref{mbac} illustrates the distribution of motorcycle deaths in various Asian countries, highlighting that Bangladesh has the highest number of fatalities in motorcycle accidents among all Asian nations.

While helmets can protect against head injuries and prevent fatalities, there is another way that can reduce these fatalities even more: emergency assistance to the injured motorcycle rider. Most of the casualties from accidents are due to poor communication between the departments in charge of rescuing and providing medical help, which results in the late arrival of medical assistance \citep{rf3}. If the victim remains unattended, it puts the life of the victim in more danger as time goes by. It is stated that a delay of 5 minutes or more to initiate the rescue operations increases the fatality rate by 10\% to 20\% \citep{rf4}. Again, the family and friends of the victim remain completely unaware of the situation when an accident occurs, as the victim is not able to communicate with them in most cases. Only family members understand how stressful it is when they are unable to contact their loved one and are unaware of his or her whereabouts. This research aims to come up with a solution that can mitigate these problems to a greater extent.

\begin{figure}[!htbp]
\centering
\includegraphics[width=0.8\textwidth]{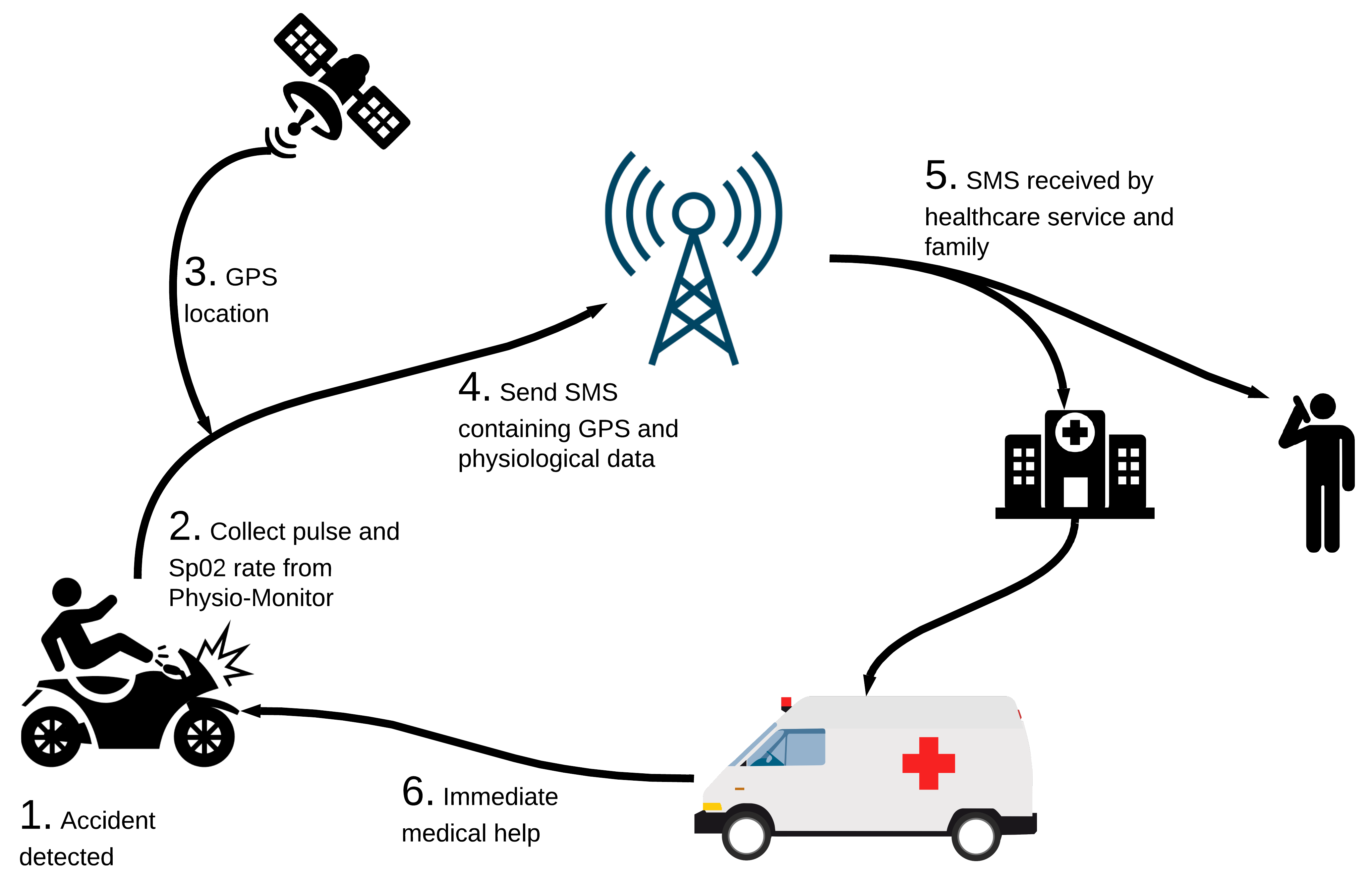}
\caption{A glimpse of the proposed system.}
\label{glimpse}
\end{figure}

This paper presents a proposed solution aimed at reducing the fatalities associated with motorcycle accidents. Our approach involves the development and implementation of an accelerometer-based accident detection system. The system is designed to promptly detect an accident occurrence and subsequently notify the victim’s friends and family members. The notifications will include detailed information regarding the accident, the current condition of the victim, and the real-time location of the injured individual shown in Fig.~\ref{glimpse}. The primary contributions of our study encompass the following:
\begin{itemize}
    \item Detecting a motorbike accident as soon as it occurs
    \item Notifying friends and family members about the condition through SMS
    \item Tracking the real-time location of the victim 
    \item Real-time tracking of physiological indicators, including heart rate and blood pressure
\end{itemize}

\section{Review of existing systems}
In the field of motorbike accident mitigation, various researchers have proposed different approaches to address the issue. \citet{rff4} focused on developing a system that utilizes a heart rate sensor to detect abnormal heartbeat patterns, thereby inferring the occurrence of an accident. Once an accident is detected, the system transmits the accident location to the nearest hospital via a Bluetooth-based Android application. Another research article by \citet{rf4} introduced the continuous utilization of a high-definition camera as a visual input mechanism. The captured raw images are processed by a Raspberry Pi, which employs image analysis techniques to detect accidents. Once an accident is identified, the system promptly notifies the relevant authorities. \citet{rff6} presented an approach where the values obtained from an accelerometer embedded in a helmet are continuously monitored. Through analysis of these accelerometer readings, accidents can be detected and emergency notifications containing the GPS location information are sent to designated contacts. In the study conducted by \citet{rff5}, a six-axis accelerometer was employed for accident detection. The accelerometer data were analyzed to identify abnormal patterns indicative of accidents. \citet{rff7} proposed a helmet equipped with an automatic safety headlight system. This system utilized an accelerometer, along with other sensors and motors, to respond to the rider's facial movements. The aim was to enhance safety by providing automatic and adaptive lighting based on the rider's actions.
\par
Alternative approaches have been explored by researchers, including the utilization of smartphones for accident detection. \citet{rf17} proposed a comprehensive system incorporating smartphones, GSM, GPS technologies, vehicular ad-hoc networks, and mobile applications for accident detection. Similarly, \citet{rf18} introduced a smartphone-based system that leverages accelerometers and acoustic data to automatically detect traffic accidents on platforms like iPhone and Google Android. The system promptly notifies a central emergency dispatch server and provides situational awareness. In a similar vein, \citet{rf19} employed smartphones' accelerometers and GPS capabilities for accident detection. They harnessed these sensors to identify sudden changes in motion and relayed the accident information along with GPS coordinates.  A different approach was taken by \citet{rf20}, where accident detection relied solely on GPS technology. By tracking a vehicle's speed, accidents were inferred based on significant drops in speed below the authorized limit, and the accident location was reported to an alert service center. \citet{rf21} proposed a system for detecting bicycle accidents that combined the MPU6050 (gyro sensor and accelerometer), SIM808 (GPS+GPRS+GSM), Raspberry Pi 3 Model B, and Arduino Uno. The system placed on the bike's surface utilized sensors to detect accidents and quickly transmitted messages containing the biker's contact information and location to nearby hospitals, law enforcement, and registered family members.
\par
\citet{rf8} introduced a smart helmet designed to prioritize air quality, helmet removal, and collision detection. The helmet incorporates sensors to detect hazardous gases such as CO, SO\textsubscript{2}, and NO\textsubscript{2}. Additionally, it employs a commercially available infrared sensor to detect helmet removal. Utilizing these sensor readings, the system calculates the Head Injury Criteria (HIC) using an accelerometer to identify accidents. \citet{rf9} proposed a Vehicle-to-Vehicle communication model that enables an affected vehicle to send warning messages to nearby vehicles and a Control Database Server (CDS). However, the system incorporates a switch that allows the signal to be terminated if the injured individual does not require immediate first aid or if the injuries are not severe. Another strategy for reducing traffic accidents involves lane detection, where a defined region of interest with high contrast, referred to as the lane area, is identified using the Hough transform. \citet{rf11} implemented this approach by selecting appropriate threshold values to designate lane boundaries. They further utilized the Partial Least Square Filtering (PLSF) and Euclidean distance formulas to refine the lane detection process. In a different vein, \citet{rf12}, developed a tracking and notification system that transmits vehicle location data to a monitoring device. The system incorporates a radio service and includes a notification system to facilitate timely alerts. A comparison of some of the literature's work, employed technology, beneficial effects, and potential drawbacks is shown in Table~\ref{table:limit}.

\begin{table}[!ht]
\begin{center}
\scriptsize
\caption{Analysis of some existing systems}
\begin{tabular}{p{0.08\linewidth}  p{0.2\linewidth}  p{0.2\linewidth}  p{0.2\linewidth} p{0.2\linewidth}} 
\toprule
\textbf{Reference} & \textbf{Proposed research work}  & \textbf{Technologies Used} & \textbf{Advantages} & \textbf{Limitations} \\
\midrule 
\citep{rf4} & Detects accidents, notifies and sends the location to the nearest hospital  & Heart rate sensor, GPS, and Bluetooth-based Android application & Good chance of getting proper treatment as soon as possible as it notifies the hospitals & There should always be an internet connection, or it cannot notify or send locations to the hospitals\\
\midrule 
\citep{rf5} & Notifies the authority after detecting the accident using the raw images & Image processing, GSM, GPS, and Raspberry Pi & Image processing algorithms are used, which make the detection more accurate & Severe weather conditions sometimes distort the image quality, which might create problems in detection\\
\midrule 
\citep{rf7} & Smart helmet controlled by a microcontroller with a Force Sensing Resistor (FSR) and a BLDC Fan for speed detection. & BLDC fan, Force Sensing Resistor (FSR), and Radio Frequency modules & Reducing the impact of collisions and preventing motorcycles from being taken.& The helmet's alarm system is activated when the speed exceeds a predetermined threshold, and the motorcycle's engine will not start until the rider has secured the helmet.\\
\midrule 
\citep{rf9} \citep{rf10}  & An accident is detected using the V2V communication system & GPS, ZigBee, and VANET & All the vehicles are interconnected with each other and the main control database server, which takes action as needed & This is only possible on a large scale, and maybe only the government or a big organization can take this initiative\\
\midrule 
\citep{rf11}  & Detect accidents using the Hough transform technique of image processing & Image processing, GPS, cameras, and Raspberry Pi & The Hough transform technique can be used only for proper road lane detection, which notifies the rider if they are driving improperly & It does not do anything once the accident has happened\\
\midrule 
\citep{rf12}  & Detects accidents using an accelerometer, a smoke sensor, and a shock sensor & GPS, GPRS, and Arduino & As there is more than one sensor, it can detect the accident more accurately & Needs more sensors to detect the accident and so it is expensive.\\
\midrule 
\citep{rf13} & Detects accidents and measures the intensity of the accidents & ATmega 328 microcontroller, Force Sensor, GSM and GPS module, Servo Motor, PI camera, etc. & IoT and AI-based systems detect accidents, take photos with a PI camera and send them to the cloud, and calculate the false alarm and accident intensity & Requirement of continual internet access\\
\midrule 
\citep{rf14} & Detects accident and notifies the rescue team & ADXL335, Arduino Uno, Vibration Sensor, GPS, GSM module, Thermistor & A safer system that can save two-thirds of the population from dangerous road accidents & The biggest problem is the blockage of signal transmission by mountains, high buildings, and tunnels\\
\bottomrule
\end{tabular}
\label{table:limit}
\end{center}
\end{table}

\section{Proposed system}
Our entire system can be divided into two sub-systems that work simultaneously. The first one is the accident detection circuit, whose main purpose is to detect the accident as soon as it occurs using an accelerometer, then derive the real-time location of the accident site through GPS, and finally send the SMS to the emergency contact number using the GSM module.
\par
The second one is the physio monitor. This is a watch-like device that can be worn on the wrist like a wristwatch. The main purpose of the physio monitor is to collect the pulse rate and oxygen saturation rate of the rider and continuously send the data to the accident detection circuit. The pulse rate and oxygen saturation rate are collected using the MAX30100 sensor, and this subsystem is controlled by a smaller version of the NodeMCU. The continuous communication between the accident detection circuit and the physio monitor is established by using the Wi-Fi protocol.

\subsection{Accident detection}

\begin{figure}[!ht]
\centering
\begin{subfigure}{0.6\textwidth}
    \centering
    \includegraphics[width=0.7\linewidth]{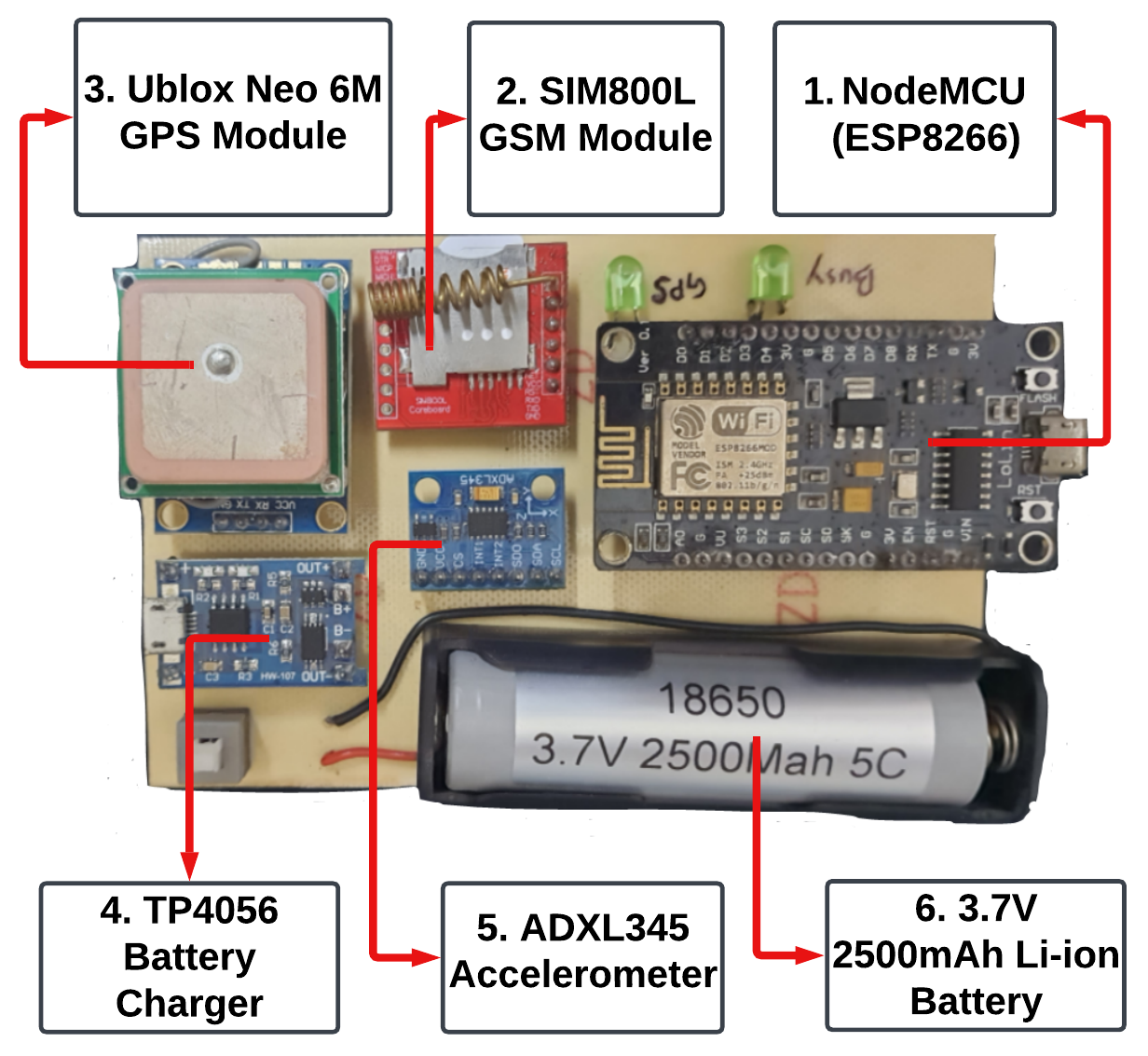}  
    \caption{}
    \label{fig:sub-second}
\end{subfigure}
\begin{subfigure}{0.35\linewidth}
    \centering
    \includegraphics[width=0.85\linewidth]{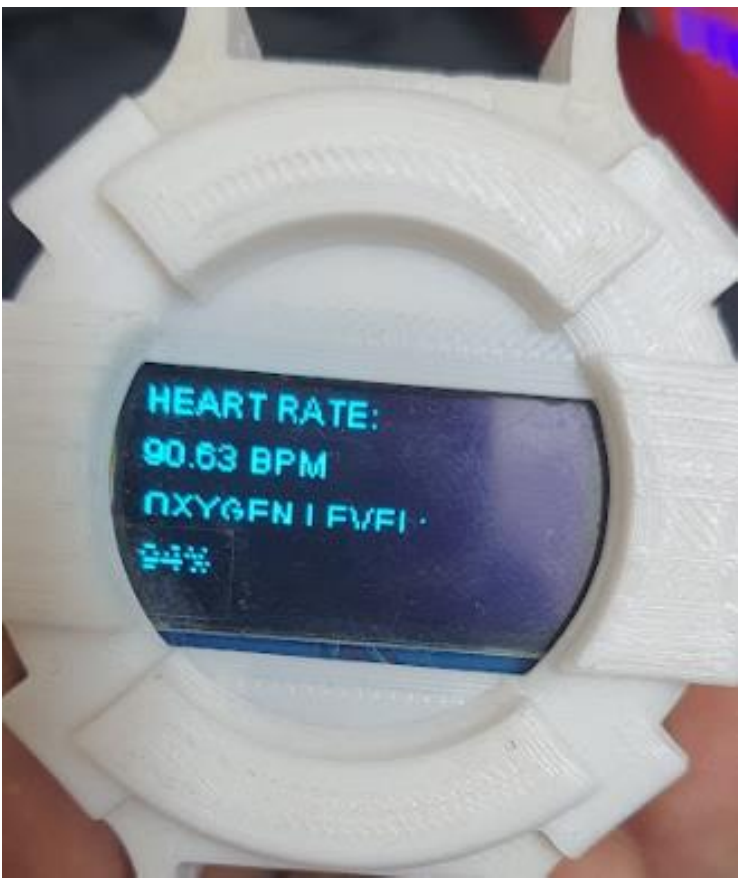}  
    \caption{}
    \label{phym-cir}
\end{subfigure}
\caption{Hardware of (a) accident detection and (b) physio-monitor circuits.}
\label{fig:fig_det_ckt}
\end{figure}

\begin{figure}[!ht]
\begin{subfigure}{1\textwidth}
  \centering
  \includegraphics[width=0.8\linewidth]{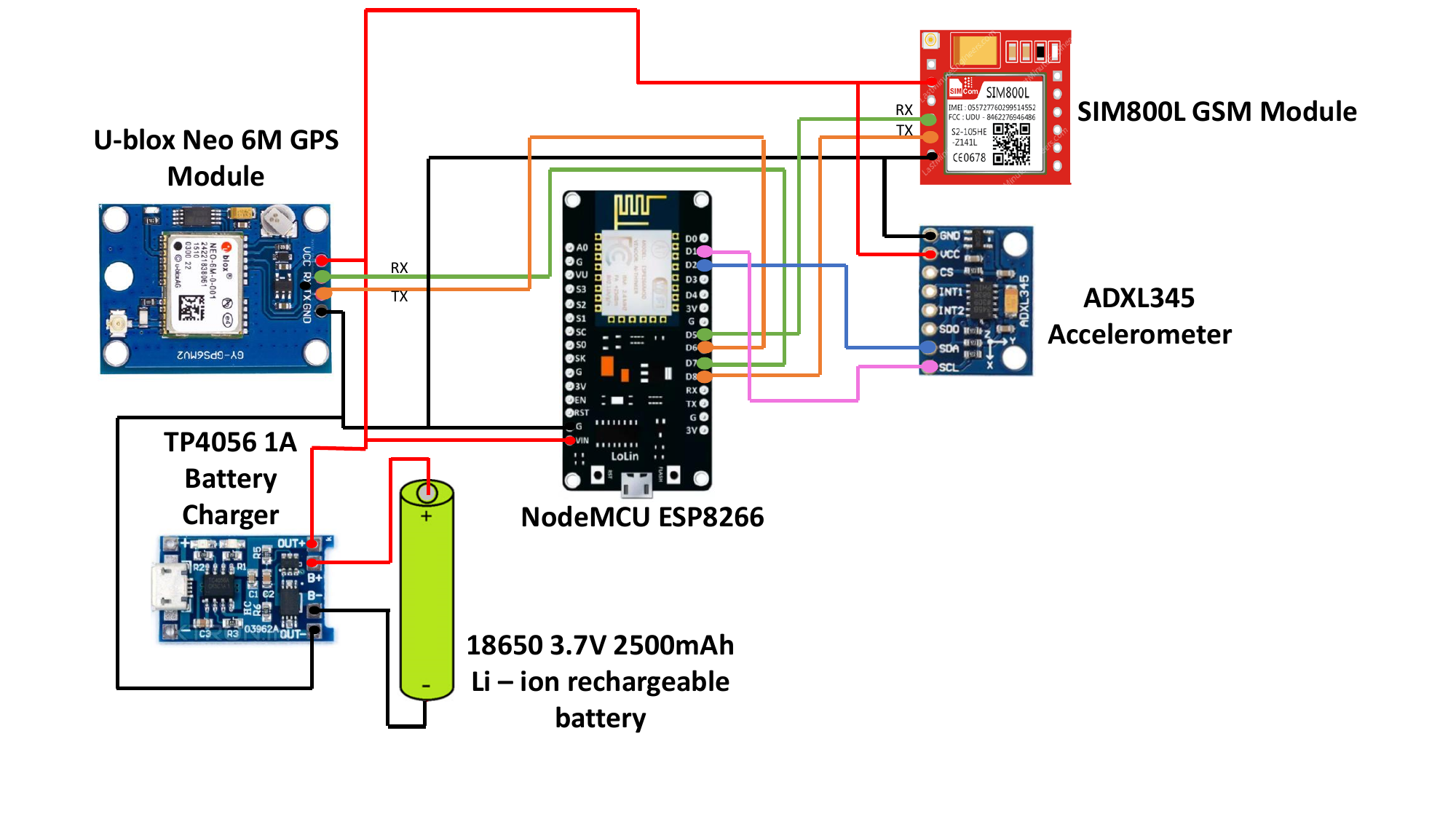}  
  \caption{}
  \label{fig:sub-second-scm}
\end{subfigure}\\
\begin{subfigure}{1\textwidth}
  \centering
  \includegraphics[width=0.8\linewidth]{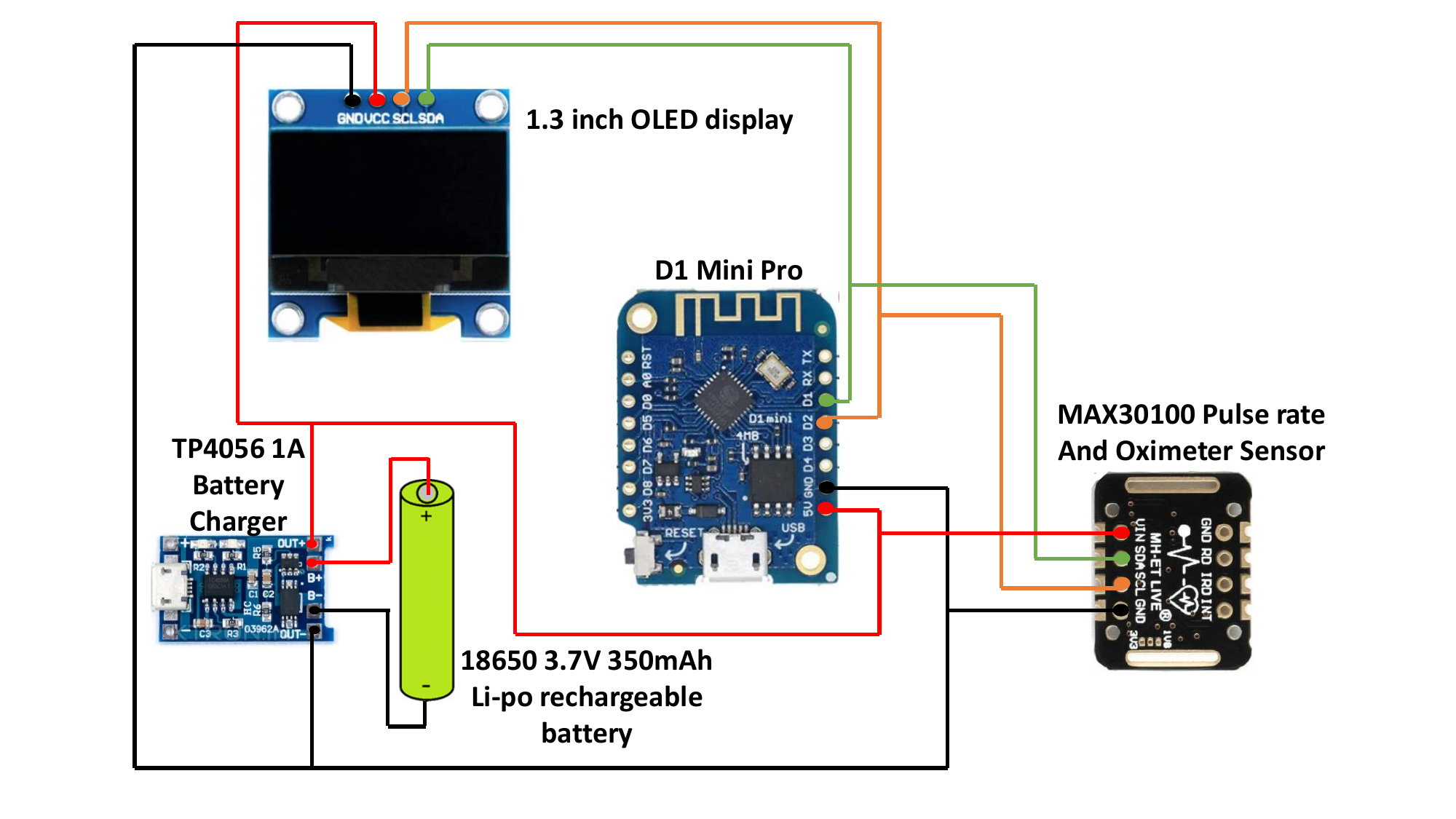}  
  \caption{}
  \label{phym-cir-scm}
\end{subfigure}
\caption{Schematic of (a) accident detection and (b) physio-monitor circuits.}
\label{fig:fig}
\end{figure}

The system's accident detection portion will be handled by an accelerometer built into the helmet's onboard circuit. In this system, we selected an ADXL345 accelerometer with digital output, shown in Figs.~\ref{phym-cir} and \ref{phym-cir-scm}, since it simplifies interfacing with the NodeMCU. The ADXL345 accelerometer can measure tilt angles along three axes, although roll and pitch angles with the x and y axes are the parameters needed to detect an accident.

We concluded that when the accelerometer reading goes below 200 and more than +200 in both the x and y axes, it reaches such an angle with the ground that it is impossible for a bike to still have a grip on the road after testing the accelerometer by placing it in a helmet. The NodeMCU receives the accelerometer reading constantly and determines that an accident has occurred when it gets a reading of less than 200 or more than 200 for more than 5 seconds.

\subsection{Pulse rate and SpO$_{2}$ saturation monitor}
The pulse rate and SpO$_{2}$ saturation meter, shown in Figs.~\ref{fig:sub-second} and \ref{fig:sub-second-scm}, will operate continuously in the hand, similar to a wristwatch. The pulse rate and SpO$_{2}$ saturation monitor transmit physiological data to the helmet continuously because the NodeMCU (ESP8266) in the helmet circuit and the D1 Mini Pro in the monitor are both wirelessly connected through Wi-Fi.

\subsection{Sending SMS to the emergency contact numbers}
The rider's current location, combined with the pulse rate and SpO$_{2}$ saturation data, which are continuously uploaded to the onboard circuit through Wi-Fi, are sent via SMS to the emergency contact number as soon as the accident is detected. The message-sending operation will be carried out by SIM800L.
\par
\begin{figure}[!htbp]
\centering
\includegraphics[width=0.7\textwidth]{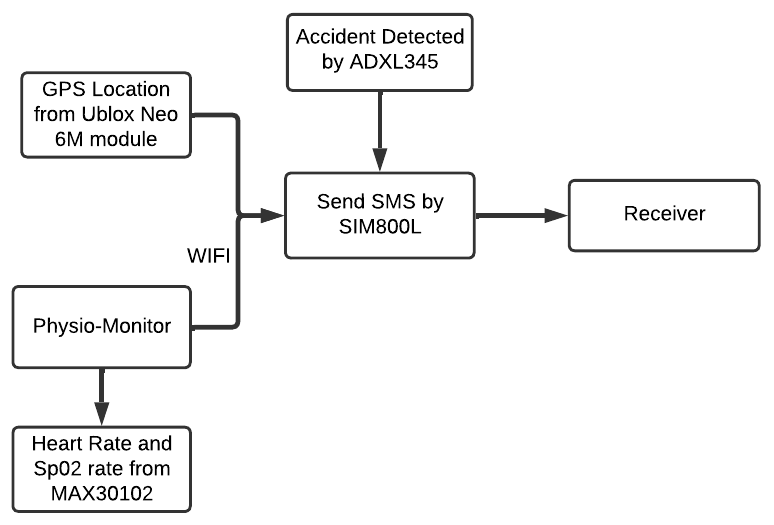}
\caption{Block diagram of the proposed system.}
\label{bikesafe}
\end{figure}

The microcontroller configures the SIM800L GSM module using AT instructions as a component of the accident detection system, enabling SMS to be transmitted in the event of an accident shown in Fig.~\ref{bikesafe}. As the central processing unit, the NodeMCU gathers data from various sensors, including the GPS module, pulse rate sensor, and SpO$_{2}$ saturation sensor from the ADXL345 accelerometer. The ADXL345 accelerometer is specifically used to calculate the tilt angle. A predetermined range of -200 to +200, or an angle range of 20° to 160°, is chosen to detect accidents. If the accelerometer readings are outside this range, the vehicle is in an unstable posture and has lost traction, which might lead to an accident. The system confirms the accident by continuously monitoring the accelerometer data for more than 5 seconds outside the permitted range.

The device also includes a physio-monitor circuit comprising a SpO$_{2}$ saturation sensor and a MAX30100 pulse rate sensor. These sensors continuously collect data and transmit it to the accident detection circuit over Wi-Fi. Furthermore, the Ublox Neo-6M GPS module can determine the rider's location. In the case of an accident, after gathering data from the GPS and physio-monitor, the microcontroller utilizes AT instructions to activate the SIM800L GSM module, which subsequently sends an SMS alert to the emergency contact numbers with crucial data regarding the rider's health.
\par
A rechargeable battery with a voltage of 3.7 V, called an 18650 Li-ion, powers the accident detection circuit. This voltage level is compatible with the needs of the majority of components, which run at 3.3 V. However, a range of 3.4 to 4.4 V is necessary for the GSM module to operate at its best. Low Dropout Regulators (LDR) are incorporated into the circuit to automatically reduce higher voltages to the required 3.3 V to fix this concern. Similarly, the physio-monitor circuit powers a rechargeable 3.7 V, 350 mAh Li-po battery. Both battery systems can be easily refilled with a Type-B USB connector and a regular 5 V converter.

\section{Multiple design approaches}
In addition to the use of accelerometers, other methods exist for the detection of accidents. Two of these alternatives are discussed, followed by a comparative analysis demonstrating the superiority and efficacy of our system.

\subsection{Detection with drowsiness detection and image processing}
Most accidents are caused by drowsiness and not reaching the hospital in time. In this approach, driver's drowsiness can be diagnosed through image processing, and they can be alerted with a buzzer. Even if an accident occurs, vibration sensors might be able to find it. All the sensors are connected to the Raspberry Pi, which further collects data from the sensors and the driver’s eyes. The victim’s location is transmitted through a GPS system, which sends a message to the victim’s parents via a GSM module.

\begin{figure*}[!ht]
\centering
\includegraphics[width=0.85\textwidth]{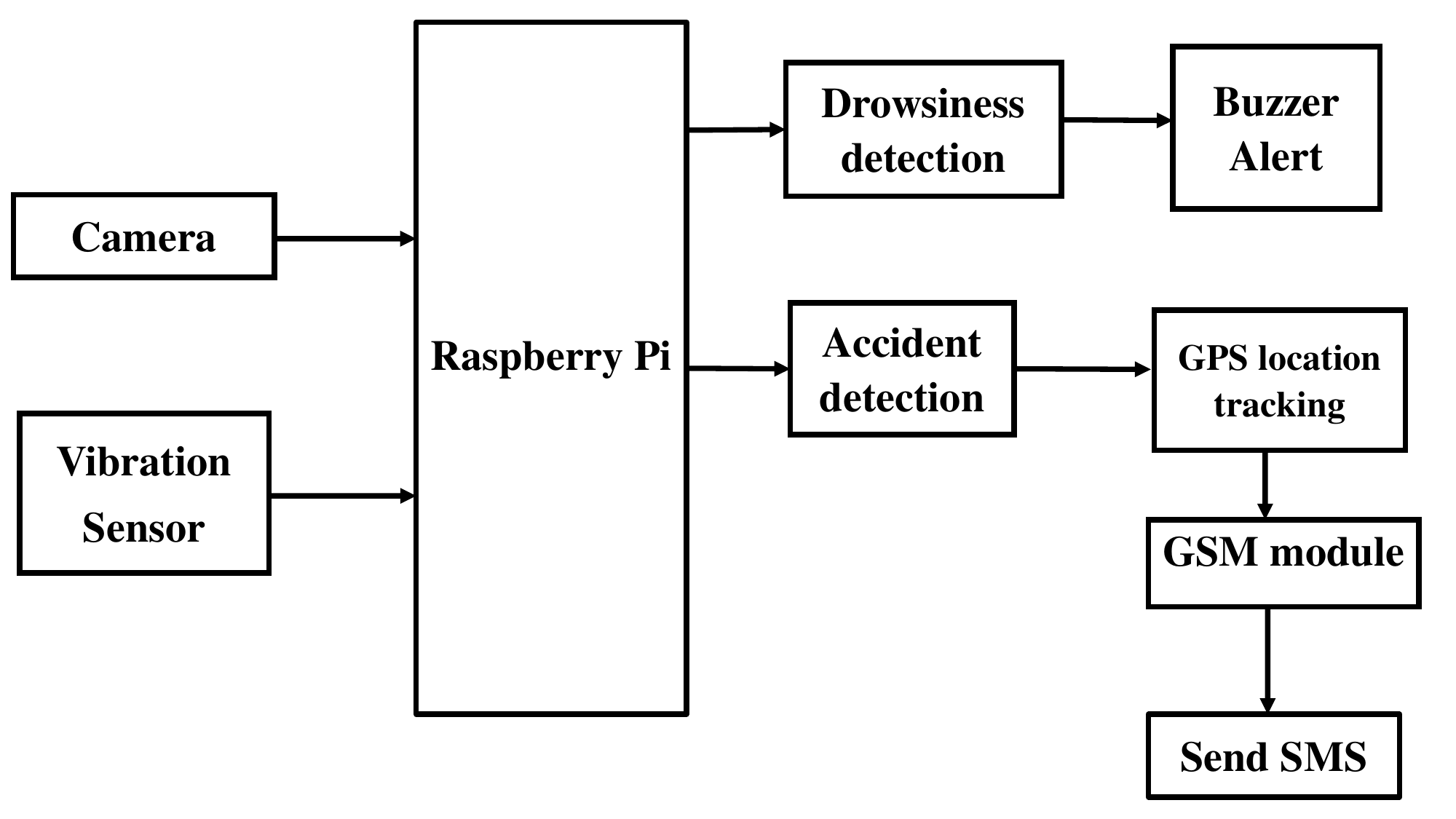}
\caption{Block diagram for drowsiness detection system-based approach.}
\label{bd_dd}
\end{figure*}

In this design, image processing will be solely used for detecting if the rider is drowsy or not. The Haar cascade method, which is a very effective way to detect the opening and closing of the eyes, will handle the image processing. Image processing requires a lot of processing memory, which is why, in this design, we have used the Raspberry Pi in order to execute all the operations. If the system detects any drowsiness by the rider, it will immediately make a sound so that the rider gets alerted about the situation. According to this design, in the event of an accident, the vibration that results from a collision will be able to detect the accident. When the system detects an accident, it will investigate how long the rider's eyes were closed. If it remains closed for more than 20 seconds, then the system will consider it an accident and notify the contacts.

\subsection{Detection using FSR}
For the enhanced safety of motorcycle riders, this accident detection approach to safety system construction necessitates a helmet and a speed alert. The pressure or force the rider experiences during an accident is measured. In the first one, we have proposed a system that contains a force-sensing resistor (FSR) to measure the pressure to detect the accident. A force-sensing resistor is a material whose resistance changes when a force, pressure, or mechanical stress is applied. They are also known as force-sensitive resistors and are sometimes referred to by the initialism FSR. They have used NodeMCU to connect the inter-system to the mobile application. Then comes the GPS module, which gets location information from satellites. This processing is done by the microcontroller and sent to the GSM modem. Finally, the GSM modem sends the information to the respective mobile numbers.
\begin{figure*}[!ht]
\centering
\includegraphics[width=0.28\textwidth]{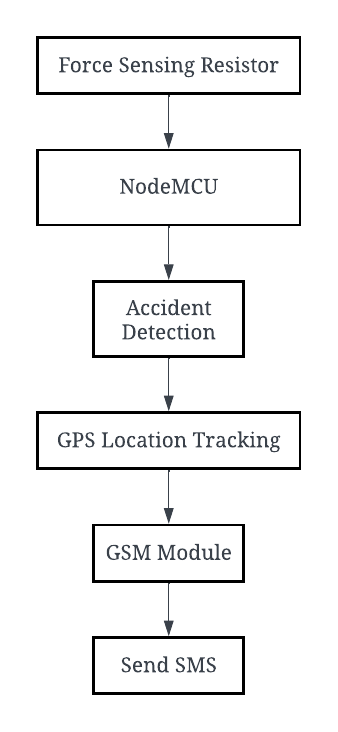}
\caption{Block diagram of FSR-based approach.}
\label{fsr}
\end{figure*}
\par
The entire device will be attached to the helmet, and when a collision occurs, whether with another vehicle or with the road surface, the FSR present in the helmet will begin to provide data on the helmet's intensity. The threshold value will be given after finding out at which value the collision or accident is actually fatal or will be severe for the rider. If the collision intensity goes beyond the threshold, the system will detect it as an accident.

\subsection{Reason for not using FSR and drowsiness detection}
\subsubsection{Power consumption}

Any portable system, including the one depicted, must be power-efficient. Because the entire system is powered by a battery, power consumption must be carefully taken into account. If FSRs were to be used for accident detection, it would take a large number of sensors, roughly 16, to effectively cover the helmet, which would increase power consumption and add bulk and complexity.

The sleepiness detector, which uses a Raspberry Pi as the microprocessor and an integrated camera, also uses a lot of electricity. The Raspberry Pi 4 uses 2.7 W of power when it is not in use and 5.1 W when it is, with an extra 0.4 W to 1.4 W used by the camera. Such power requirements make it difficult for portable gadgets to operate efficiently and call for larger batteries, which further leads to an increase in weight.

The suggested solution, in contrast, exhibits greater power efficiency. As shown in Table~\ref{tab:compv}, similar input voltage ranges are supported by the ADXL345 accelerometer, NodeMCU, Ublox Neo 6M GPS module, and SIM800L GSM module, enabling efficient power delivery from a 3.7 V Li-ion battery. Additionally, as compared to other systems, their current consumption is rather low, improving total power efficiency.
In conclusion, the suggested solution is more power-efficient than competing ideas, making it a better option for portable applications.

\begin{table}[!ht]
\begin{center}
\scriptsize
\caption{Voltage and current consumption for different components}
\label{tab:compv}
\begin{tabular}{ccc} 
\toprule
\textbf{Component} & \textbf{Input Voltage (V)} & \textbf{Current Consumption (mA)} \\
\midrule  
NodeMCU (ESP8266) & 3.3 & 70\\
ADXL345 & 3.3 & 0.15\\
Ublox Neo 6M GPS Module & 2.7$–$3.6 & 45\\
SIM800L GSM Module & 3.4$–$4.4 & 18 (standby)\\
\bottomrule 
\end{tabular}
\end{center}

\end{table}

\subsubsection{Exposure to the outer environment}
Both systems, which comprise FSR and image processing methods of accident detection, require the system to be exposed to the outer environment. For example, the FSRs should be attached to the surface of the helmet so that it can detect the force of the collision. On the other hand, the camera will also be exposed so that it can detect drowsiness as well as if the eyes have been closed. Long-term outside exposure can cause damage to the components due to humidity, dust, and heat, which increases the possibility that the outputs will be compromised.
\par
On the other hand, our system does not need any type of external exposure. All the components can be placed inside the helmet, where they will be protected from any type of outside factor that could cause damage to the system.

\subsubsection{Maintainability}
Every system requires maintenance in order to function properly and reliably. Since this is self-evident, the next question is how frequently that maintenance is required, how simple it is to perform, and whether all of the components are always available. If we use FSR, it is obvious that it will suffer frequent damage because it will be constantly exposed to the outside environment. What’s more concerning is that there will be around 16 pieces of FSR attached to the helmet. It is very possible that one or two might get damaged. Running a proper diagnostic to find the damaged ones will take unnecessary time. The components used in this system are widely available on the market, so obtaining them will be simple.
\par
Again, the camera of the drowsiness detector would be exposed, and it might need maintenance, but not as frequently as FSR. Dust, humidity, and water can always create problems and damage it. If it gets damaged, there is no other way to change it, which increases the price of maintenance. If we compare, as our system is totally protected inside a helmet and each of the components is easily accessible, it will require less maintenance.

\subsubsection{Budget comparison}
As of 2021, an analysis of the prices for the components necessary to build different systems reveals noteworthy findings. Among the options, the ADXL345 accelerometer-based detection system emerges as the most cost-effective, with a price of $3,130$ BDT. On the other hand, the systems involving image processing for drowsiness detection and force-sensitive resistors (FSR) command higher price tags, with respective costs of $9,650$ BDT and $14,628$ BDT. This information highlights the varying expenses associated with different detection systems, and it provides valuable insights for researchers and practitioners seeking to make informed decisions regarding the selection and budgeting of components for their projects.
The budget comparison is summarized in \autoref{tab:comparison}, and the itemized budgets for each solution are presented in Tables~\ref{table:compprice}, \ref{tab:compprice1} and \ref{tab:compprice2} in the appendix.

\begin{table}[!ht]
\begin{center}
\scriptsize
\caption{Comparative analysis of accident detection techniques}
\label{tab:comparison}
\begin{tabular}{p{2cm}p{1.5cm}p{1.5cm}p{1.45cm}p{1.0cm}p{1.8cm}p{1.8cm}p{1.8cm}}
\toprule
\textbf{Design} & \textbf{Data Transfer Rate}  & \textbf{Remarks} &\textbf{Usability}&\textbf{Budget (BDT)}&\textbf{Maintenance}&\textbf{Accessibility}&
\textbf{Availability}\\
\midrule  
\textbf{Approach-01 (Accelerometer)}
&85kbps
At 850-1900 MHz
&Provide maximum performance&Convenient&3130&Simple&Satisfied&Mostly Available\\
\midrule 
\textbf{Approach-02 (FSR)}
&25mbps
At 2.4 GHz
&Difficult as too many sensors are needed&Inconvenient&14268&Complex&Partially Satisfied&Mostly Available\\
\midrule 

\textbf{Approach-03 (Image Processing)}
&85kbps
At 850-1900 MHz
&Does not fully satisfy our objective&Convenient&9650&Complex&Unsatisfied&Unavailable\\
\bottomrule

\end{tabular}
\end{center}
\end{table}

\section{Results and discussion}
As mentioned earlier, we primarily aim to detect an accident and send an SMS to the emergency contact number instantly, which will contain the real-time location and physiological activity data.
\subsection{Experimental results}
\begin{figure*}[!ht]
\centering
\includegraphics[width=0.7\textwidth]{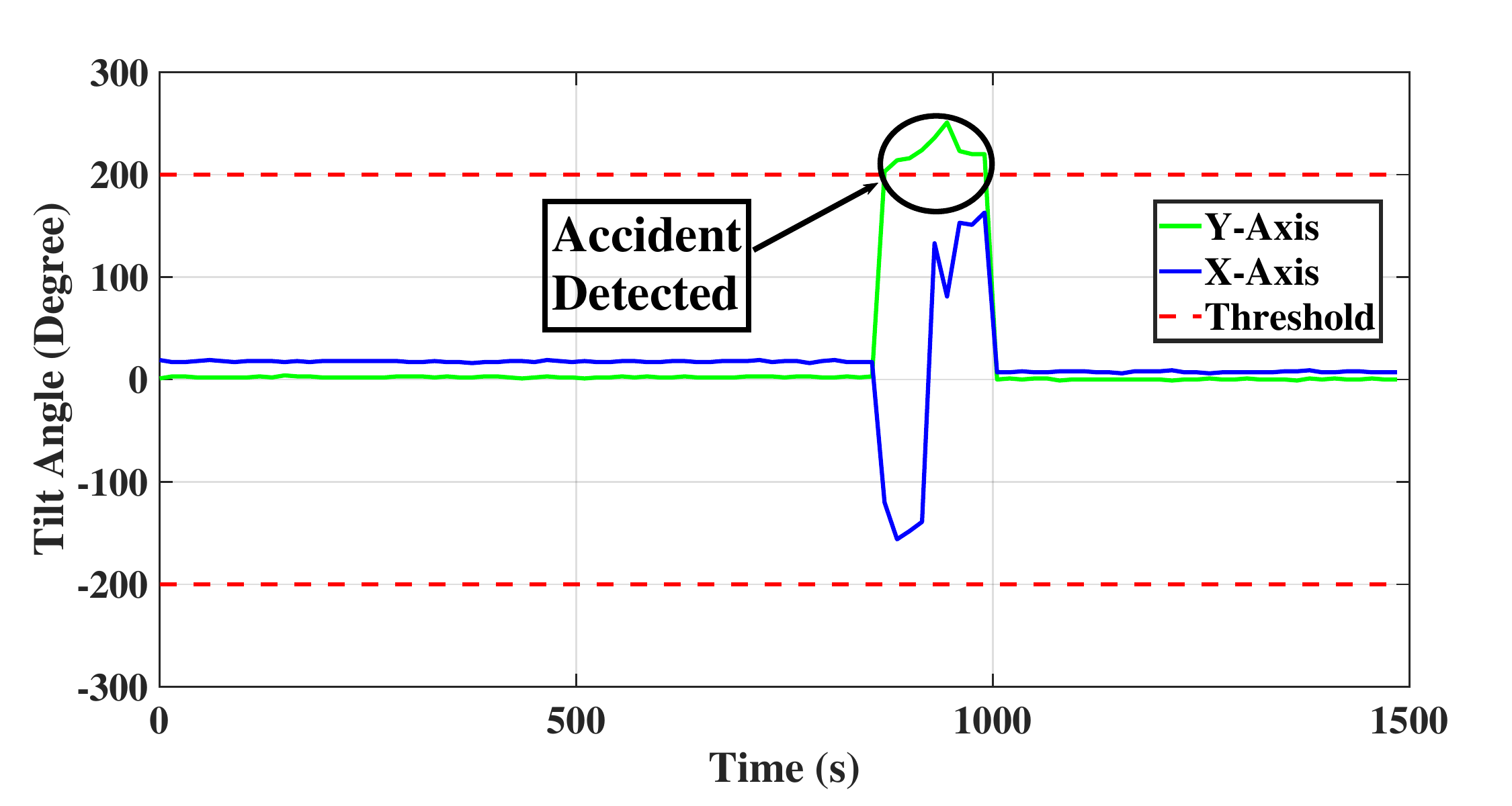}
\caption{Accelerometer reading when an accident occurs.}
\label{acclmtr}
\end{figure*}

\begin{figure}[!t]
\begin{subfigure}{0.35\textwidth}
  \centering
  \includegraphics[width=1\linewidth]{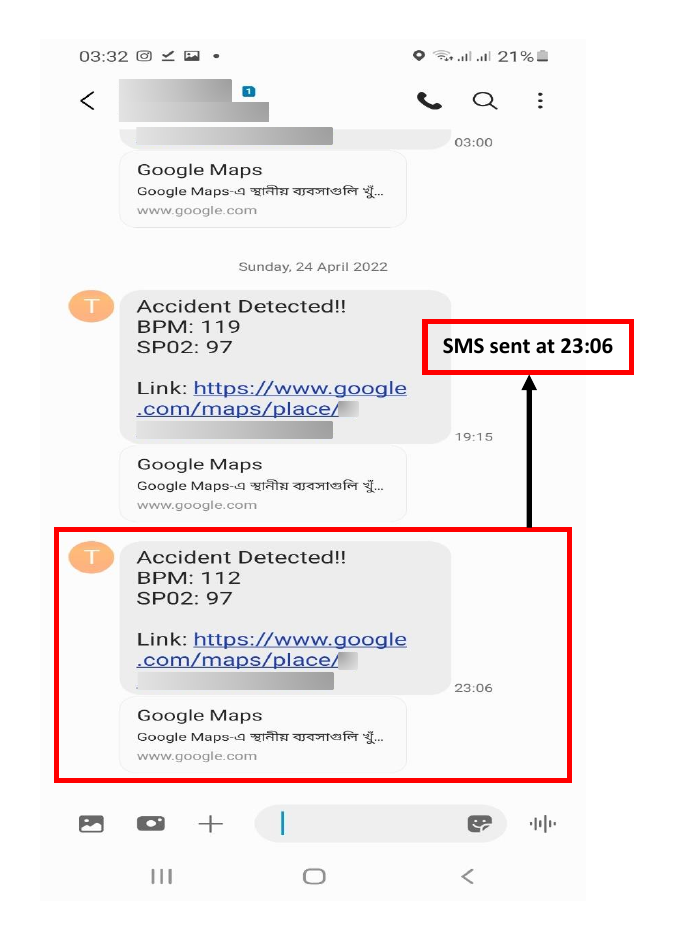}  
  \caption{}
  \label{Sms_res}
\end{subfigure}
\begin{subfigure}{0.5\textwidth}
  \centering
  \includegraphics[width=1.3\linewidth]{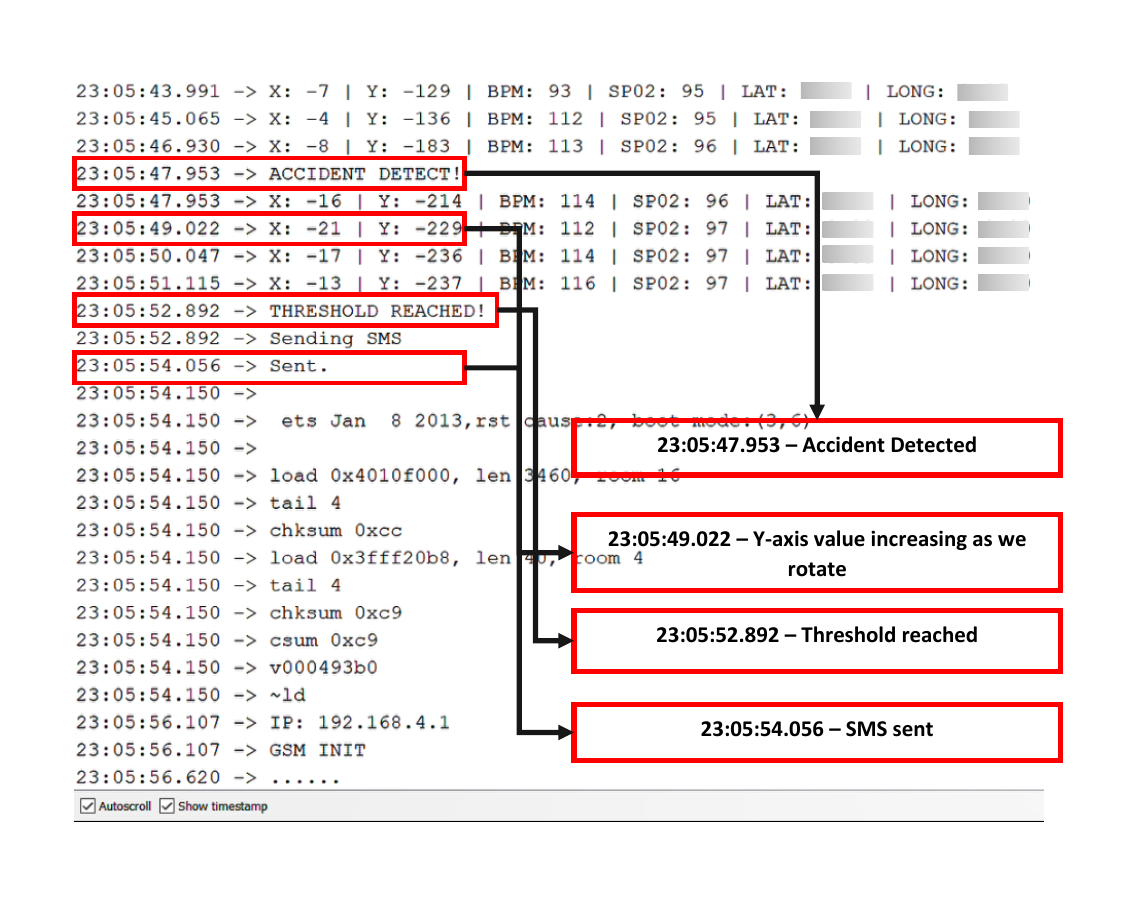}  
  \caption{}
  \label{Srl_mon}
\end{subfigure}
\caption{(a) SMS Result. (b) Serial Monitor Result.}
\label{fig:fig_sms_res}
\end{figure}

Fig.~\ref{fig:fig_sms_res} shows all the data the entire system is providing. It shows the accelerometer data for both the X and Y axes. As the accelerometer is moved towards the Y-axis, the data has continued to increase.

At 23:05:47, our system alerted us that an accident had been detected when the Y-axis value exceeded $-200$. But still, it would not send the SMS because we wanted it to check if the data was still greater than 200 after 5 seconds. As we can see, the system still recorded the values of the Y-axis around $-200$. For this reason, exactly after 5 seconds at 23:05:52, the system has determined the threshold has been reached, and thus, it will consider that an accident has surely happened. The system then informs us that an SMS is being sent and when it is complete.

Next to the values of the accelerometer are the values provided by the pulse rate monitor. The system is continuously recording data for pulse rate and oxygen saturation. Along with it, we can also see the latitude and longitude values. The pulse rate and Sp02 saturation values will be sent with the SMS directly, but the latitude and longitude will be added with the Google Maps link, which will give us the real-time location of the rider.
At 23:05:52, the system showed us that an accident had happened and an SMS had been sent. As we can see in Fig.~\ref{Srl_mon}, we received a text message from the system on our phone at exactly 23:06.

\subsection{Integration capability and economic viability}

A method of accident detection that is beneficial to the environment has already been created in several countries, such as Sri Lanka and China \citep{rf15}.
However, it is very difficult for nations with overpopulated cities, such as Bangladesh and India. India's weak accident detection system infrastructure is a major failure. Countries like Bangladesh or India should place greater emphasis on appropriately collecting data to prevent accidents and imposing strict laws.
The proposed prototype can take the place of conventional methods for detecting accidents. According to their needs, municipal governments can create a system to monitor these intelligent systems. 

The proposed accident detection and notification system is designed with reduced cost. The proposed smart helmet prototype can be constructed for a total of only USD 31.02 (BDT 3130). Refer to Table~\ref{table:compprice} in the appendix for the cost analysis. 

\section{Conclusion}
Motorcycle accidents have become one of the most common occurrences in modern society. The number of accidents has increased dramatically in recent years due to a lack of adequate protection systems and safety measures. These incidents result in numerous injuries and even fatalities. In order to mitigate this fatality problem, researchers are persistently establishing a reliable accident detection and notification system. In this paper, we propose a system for the automatic detection and notification of motorbike incidents that offers a comprehensive solution to the rising risks associated with motorcycle riding as a result of urbanization and rising living standards. By incorporating both a detection circuit and a physio-monitoring circuit, the system seeks to improve safety by promptly detecting accidents and ensuring prompt medical care. The detecting circuit of the motorcyclist's headgear consists of an ESP-8266 microcontroller, an ADXL345 accelerometer, a Ublox Neo-6M GPS module, and a SIM800L GSM module. These components detect accidents and alert emergency personnel. The accelerometer aids in detecting collisions, acceleration changes, and extreme angles that result in the motorcycle losing traction. Pulse rate and oxygen saturation sensors from the MAX30100 augment the physio-monitoring circuit's detecting circuit. This circuit measures a cyclist's pulse rate and SpO$_{2}$ saturation. The D1 Mini Pro microcontroller displays and transmits data from the physio-monitoring circuit to the detection circuit over Wi-Fi. 
The proposed system has been compared and analyzed against other systems, and its effectiveness and efficiency have been evaluated based on a variety of factors, such as cost, maintainability, accessibility, data transfer rate, and more. Experiments demonstrate that the proposed system detects incidents and notifies emergency contacts. By providing accident and physiological data in real-time, it enables first responders to administer aid rapidly. The primary objective of the system is to save lives by providing rapid and effective response mechanisms in motorbike accidents, which are especially fatal due to the riders' vulnerability and absence of protective gear. In conclusion, the proposed automatic detection and notification system for motorcycle accidents detects incidents promptly and notifies emergency contacts. The detection circuit and physio-monitoring circuit of the system seek to save lives and reduce injuries by addressing the urgent need for improved motorcycle accident safety.

\section*{Acknowledgement}
The authors would like to thank Ms Shahed-E-Zumrat, ex-lecturer at BRAC University, for her suggestions during the initial phase of the project.

\bibliography{references}

\newpage

\setcounter{section}{0}
\renewcommand{\thesection}{A\arabic{section}}

\section*{Appendix}
\section{Itemized budget of multiple design approaches to detect and notify motorbike accidents}
The cost of the power supply is not shown in the cost calculation due to integrated power chips. Supplies and microcontrollers are widely accessible and reasonably priced. Therefore, the widespread production of this ingenious accident system will maintain a fair price. The suggested accident detection method also reduces the pointless use of staff. In conclusion, the suggested accident detection strategy is thought to be less expensive than the methods now used.

\setcounter{table}{0}
\renewcommand{\thetable}{A\arabic{table}}

\subsection{Proposed system}
Table~\ref{table:compprice} presents the list of components, their quantity and price in BDT (Bangladeshi Taka) for our proposed system.

\begin{table}[!ht]
\centering
\caption{Price for different components used for the implementation of accident detection with accelerometer.}
\label{table:compprice}
\begin{tabular}{cccc} 
\toprule
\textbf{SL No.} & \textbf{Components}  & \textbf{Quantity} & \textbf{Price (BDT)} \\
\midrule  
1 & NodeMCU (ESP8266) & 1 & 420\\
2 & ADXL345 &1 & 340\\
3 & SIM800l GSM Module& 1 & 300\\
4 & Ublox Neo 6M GPS Module& 1 & 700\\
5 & TP4056 Charging Module& 2 & 50\\
6 & 18650 3.7V 2500 mAh Li-ion battery& 1 & 40\\
7 & MAX30100 Sensor & 1 & 280\\
8 & 1.3inch OLED display & 1 & 450\\
9 & D1 Mini Pro & 1 & 280\\
10 & 3.7V 350 mAh Li-po battery & 1 & 270\\\midrule 
 & \textbf{Total} &  & \textbf{3,130}\\
\bottomrule
\end{tabular}
\end{table}

\subsection{Other possible solutions}
Table~\ref{tab:compprice1} and Table~\ref{tab:compprice2} present the list of components, their quantity and price in BDT (Bangladeshi Taka) for alternative solutions including FSR and drowsiness detector, respectively.

\begin{table}[!t]
\centering
\caption{Price for different components used for accident detection system using FSR.}
\label{tab:compprice1}
\begin{tabular}{cccc} 
\toprule
\textbf{SL No.} & \textbf{Components}  & \textbf{Quantity} & \textbf{Price (BDT)} \\
\midrule  
1 & Arduino UNO & 1 & 620\\
2 & Force Sensing Resistor &16 & 9,920\\
3 & SIM800l GSM Module& 1 & 400\\
4 & Ublox Neo 6M GPS Module& 1 & 800\\
5 & nRF24L01 $+$ 2.4GHz RF Transceiver Module& 2 & 240\\
6 & Arduino Micro & 1 & 650\\
7 & Pulse Rate Sensor & 1 & 760\\
8 & ESP8266 NodeMCU Wi-Fi Module Lua V3 & 1 & 528\\
9 & 0.96 OLED Display & 1 & 350\\\midrule 
 & \textbf{Total} &  & \textbf{14,268}\\
\bottomrule
\end{tabular}
\end{table}

\begin{table}[!t]
\centering
\caption{Price for different components used for accident detection system using drowsiness detector.}
\label{tab:compprice2}
\begin{tabular}{cccc} 
\toprule
\textbf{SL No.} & \textbf{Components}  & \textbf{Quantity} & \textbf{Price (BDT)} \\ \midrule
1 & Raspberry Pi 4 (2GB) & 1 & 6,000\\
2 & Raspberry Pi Camera & 1 & 750\\
3 & Percussion Piezoelectric Vibration Sensor Module & 1 & 500\\
4 & SIM900A GSM Module & 1 & 1,600\\
5 & Ublox Neo-6M GPS Module & 1 & 800\\ \midrule
& \textbf{Total} &  & \textbf{9,650}\\
\bottomrule
\end{tabular}
\end{table}

\end{document}